\documentclass[twocolumn,showpacs,floatfix,prl]{revtex4}

\usepackage{float}
\usepackage{graphicx}

\newcommand{\bk}{{\bf k}}

\newcommand{\beq}{\begin{eqnarray}}
\newcommand{\eeq}{\end{eqnarray}}
\newcommand{\beqq}{\begin{eqnarray*}}
\newcommand{\eeqq}{\end{eqnarray*}}

\begin{document}

\title{Topological Superconductivity and Surface Andreev Bound States in Doped Semiconductors: Application to Cu$_x$Bi$_2$Se$_3$}

\author{Timothy H. Hsieh}
\affiliation{Department of Physics, Massachusetts Institute of Technology, Cambridge, MA 02139}

\author{Liang Fu}
\affiliation{Department of Physics, Harvard University, Cambridge, MA 02138}


\begin{abstract}

The recently discovered superconductor Cu$_x$Bi$_2$Se$_3$\cite{cava} is a candidate for three-dimensional time-reversal-invariant topological superconductors\cite{fuberg}, which are predicted to have robust surface Andreev bound states hosting massless Majorana fermions.  In this work, we present an analytical and numerical study of the surface Andreev bound state wavefunction and dispersion. We find the topologically protected Majorana fermions at $k=0$, as well as a new type of surface Andreev bound states at finite $k$.  We relate our results to a recent point-contact spectroscopy experiment\cite{ando}.   

\end{abstract}

\maketitle

The discovery of topological insulators 
 has generated much interest in not only understanding their properties and potential applications to spintronics and thermoelectrics but also 
searching for related topological phases in new directions.  A particularly exciting avenue is topological superconductivity\cite{fuberg, ludwig, kitaev, read, roy, zhangtsc, volovik, sato, yip, moore, nagaosa}, 
in which unconventional pairing symmetries generate nontrivial superconducting gaps in a similar way that spin-orbit coupling generates inverted band gaps for topological insulators.  
The hallmark of a topological superconductor is the existence of gapless surface Andreev bound states (SABS) hosting itinerant neutral Bogoliubov quasiparticles, which are the analog of massless Majorana fermions in high energy physics. 

There is currently an intensive search for topological superconductors.  
In particular, a recently discovered superconductor Cu$_x$Bi$_2$Se$_3$ with $T_c \sim 3K$ has attracted much attention\cite{cava}.  
It was proposed that the strong spin-orbit coupled band structure of Cu$_x$Bi$_2$Se$_3$ may favor an unconventional 
odd-parity pairing symmetry, which naturally leads to a  time-reversal-invariant topological superconductor\cite{fuberg}.  Subsequently, many experimental and 
theoretical efforts\cite{wray, heat, sample, magnetization, haolee} have been made towards understanding superconductivity in Cu$_x$Bi$_2$Se$_3$. 
In a very recent point-contact spectroscopy experiment, Sasaki {\it et al.}\cite{ando} have 
observed a zero-bias conductance peak which is attributed to SABS and seems to signify unconventional pairing\cite{abs}.

Motivated by this finding, in this Letter we study the phase diagram of odd-parity topological superconductivity and the resulting surface Andreev bound states in 
doped semiconductors with strong spin-orbit coupling, of which Cu$_x$Bi$_2$Se$_3$ is a prime candidate. 
We start from a $k\cdot p$ Hamiltonian which captures the essential features of its band structure near the Fermi surface.  
 By studying the phase diagram of the $k\cdot p$ Hamiltonian as a function of band gap, pairing potential, and doping,  
we establish three gapped phases: topological superconductor (TSC),  
topological insulator (TI), and normal band insulator (BI).  We characterize these phases in a unified way by introducing a  
topological invariant---a generalized mirror Chern number.   
We find that the odd-parity topological superconductivity in both doped BI and TI gives rise to 
surface Majorana fermions with a linear dispersion at $\bk=0$, as expected from the bulk-boundary correspondence. 
However, the SABS of the two become quite different at large $\bk$. 
In particular, we infer from the mirror Chern number that the SABS  in a superconducting doped TI  must become gapless again   near the Fermi momentum. 
This results in a new type of zero-energy SABS. 
To support these findings, we construct a two-orbital tight-binding model to calculate the SABS dispersion numerically.  
Finally we relate these results to the recent experiment\cite{ando}. 

\begin{figure}
\centering
\includegraphics[width=3in]{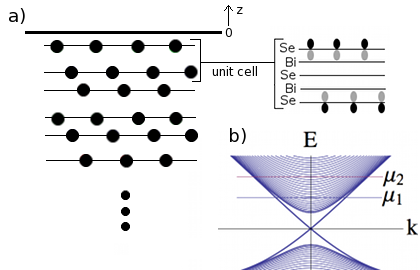}
\caption{a) Side view of a semi-infinite crystal of Bi$_{2}$Se$_{3}$. The two relevant $p_z$ orbitals are shown in the zoom-in view of the QL unit cell. 
b) Bulk and surface bands of the tight-binding model for Bi$_{2}$Se$_{3}$.  $\mu_1$ and $\mu_2$ denote two chemical potentials where the surface states have, respectively, not merged and merged into the bulk bands.}
\end{figure}

{\bf Band Structure:} 
We begin by reviewing the crystal structure of Bi$_{2}$Se$_{3}$ and the $k\cdot p$ Hamiltonian for its band structure.  
Bi$_{2}$Se$_{3}$ is a rhombohedral crystal in which each unit cell is a quintuple layer (QL) of the form Se(A)-Bi(B)-Se(C)-Bi(A)-Se(B). Here A, B, C denote three triangular lattices stacked with offset along the (111) direction, which we will define as the $z$-axis.  
Undoped Bi$_{2}$Se$_{3}$ is a topological insulator. Cu doping introduces electron carriers into the conduction band with a density of $\sim 10^{20}$cm$^{-3}$, which 
results in a small 3D Fermi surface centered at $\Gamma$. The band structure near $\Gamma$ is described by a $k\cdot p$ Hamiltonian\cite{fuberg, liu}:
\beq
H(\bk)= m \sigma_x + v_z k_z \sigma_y + v \sigma_z (k_x s_y - k_y s_x). \label{kp}
\eeq
Here $\sigma_z=\pm1$ labels the two Wannier functions which are primarily Se and Bi $p_z$ orbitals on 
the upper and lower part of the QL unit cell respectively (see Fig.1), and each has a two-fold spin degeneracy labeled by $s_z=\pm 1$.   
As explained in Ref.\cite{fuberg}, the form of $H(\bk)$ can be deduced entirely by time reversal and crystal symmetry considerations. 
The physical origin of various terms in (\ref{kp}) will become clear from an explicit tight-binding Hamiltonian later. 
While the magnitudes of the parameters $m, v_z$ and $v$ in $H(\bk)$ have been obtained by fitting the band dispersion with the angle-resolved photoemission spectroscopy measurement\cite{wray}, their {\it signs}  turn out to be particularly important in this work. 

{\bf Phase Diagram:} Now consider the superconducting Cu$_x$Bi$_{2}$Se$_{3}$ with the spin triplet, orbital singlet pairing proposed in Ref.\cite{fuberg}. 
The mean-field Hamiltonian is given by
\beq
H_{\rm MF}& =& \int d \bk [c_{\bk}^\dagger, \bar{c}_{-\bk}] H_{\rm BdG} (\bk)
\left[
\begin{array}{c}
c_{\bk} \\
\bar{c}_{-\bk}^\dagger
\end{array}
\right], \nonumber \\
c_{\bk}^\dagger &=& (c^\dagger_{\bk\uparrow}, c^\dagger_{\bk\downarrow}), \; \bar{c}_{\bk} = (c_{\bk\downarrow}, -c_{\bk\uparrow})\nonumber \\
H_{\rm BdG}(\bk)& =& ( H(\bk) - \mu) \tau_z  + \Delta \sigma_y s_z \tau_x. \label{bdg}
\eeq
Here $H_{\rm BdG}$ is the Bogoliubov-de Gennes (BdG) Hamiltonian; $\tau_{x, z}$ are Pauli matrices in Nambu space; 
$\Delta$ is the pairing potential; $\mu$ is chemical potential. 

The Hamiltonian (\ref{bdg}) exhibits three topologically distinct gapped phases as a function of the band gap, pairing potential and doping. 
At zero doping ($\mu=0$) and in the absence of superconductivity ($\Delta=0$), 
the system is either an ordinary BI or a TI, depending on the sign of $m$.  
At finite electron doping, the chemical potential lies inside the conduction band: $\mu>0$. 
When the odd-parity pairing $\Delta$ occurs in such a doped BI or TI, the system becomes a fully gapped TSC. 
For the sake of our argument, it is useful to note that the TSC phase is adiabatically connected to the $\mu=0$ and $\Delta > |m|$ limit. 
Fig.2 shows the BI, TI and TSC phases in the $\mu=0$ phase diagram as a function of $m$ and $\Delta$.  
The topological phase transition between TSC and BI/TI occurs at $\Delta=\pm m$. 

\begin{figure}
\centering
\includegraphics[width=3in]{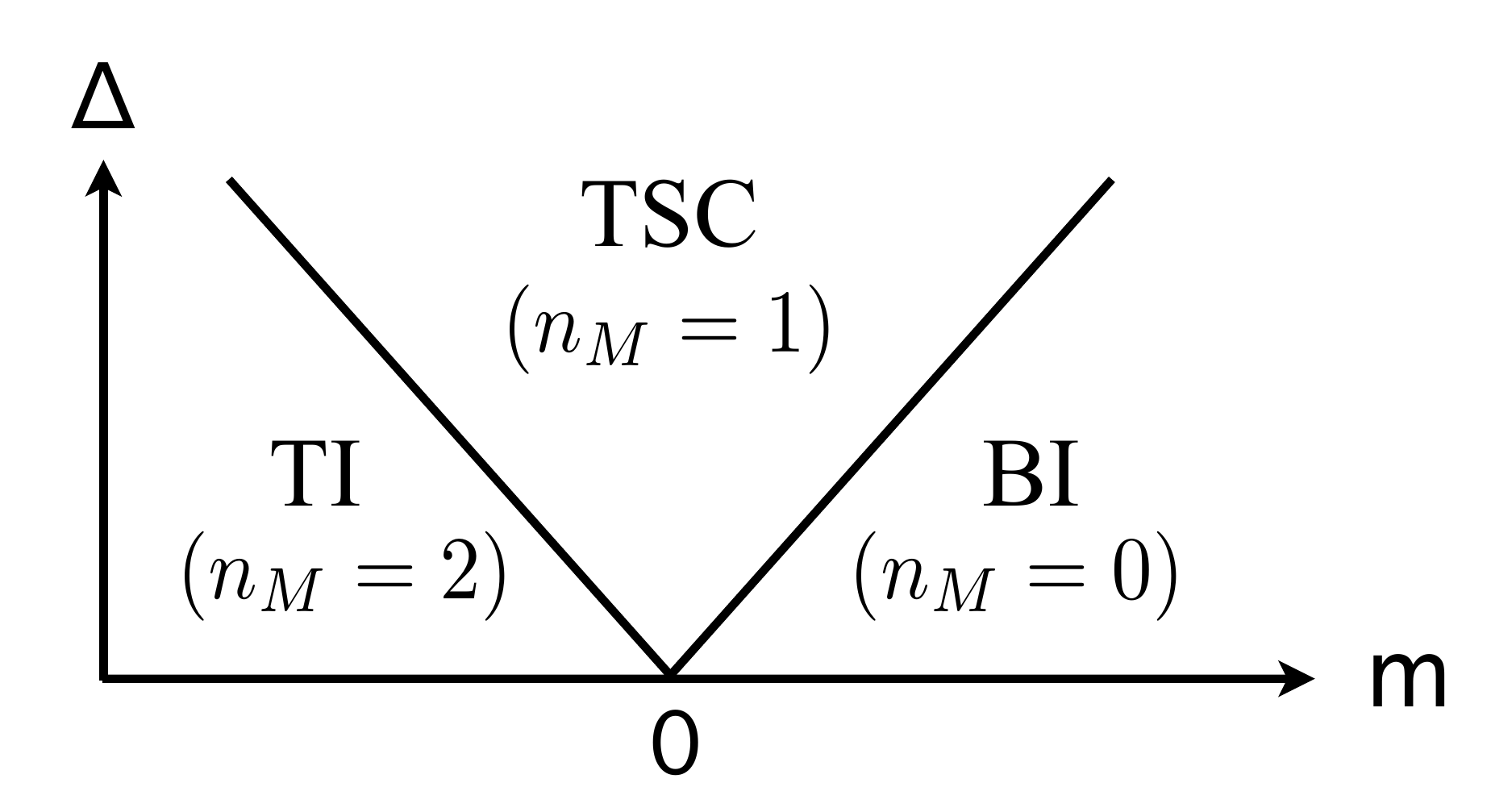}
\caption{Phase diagram of fully-gapped odd-parity superconductivity in doped semiconductors as a function of band gap $m$ and pairing potential $\Delta$, showing three gapped phases: band insulator, topological insulator and topological superconductor. 
They are topologically distinguished by the mirror Chern numbers $n_M$. }
\end{figure}

To characterize these three phases in a unified way, we introduce a common topological invariant---a generalized ``mirror Chern number'' $n_M$. Recall that reflection $M$ with respect to the $yz$ mirror plane is a symmetry element of the Bi$_2$Se$_3$ crystal point group $D_{3d}$. 
Under mirror reflection,  the band structure (\ref{kp}) is invariant, whereas the pairing order parameter in (\ref{bdg}) changes sign. 
(Mirror reflection flips $s_y$ and $s_z$.)
This sign change can be compensated by a gauge transformation $\Delta \rightarrow -\Delta$, so that the BdG Hamiltonian is 
invariant under a generalized mirror operation $\tilde{M} = M \tau_z, M=-i s_x$: 
\beq
H_{\rm BdG}(k_x, k_y, k_z) = \tilde{M} H_{\rm BdG}(-k_x, k_y, k_z) \tilde{M}^{-1}
\eeq
In particular, since $H_{\rm BdG}$ in the $k_x=0$ plane  commutes with $\tilde{M}$, a mirror Chern number $n_M$ can be defined in the same way as for an insulator\cite{teofukane}. Specifically, $h\equiv H_{\rm BdG}(k_x=0)$ is a direct sum of two subsystems 
$h_{\pm}$ with mirror eigenvalues $\pm i$ respectively:  $h_\pm = P_{\pm} h $, where $P_\pm \equiv (1 \mp i \tilde{M})/2$. 
The Chern number $n_\pm$ for $h_\pm$ satisfies $n_+ + n_- = 0$ required by time reversal symmetry. However, the difference 
defines the mirror Chern number:
$
n_M\equiv (n_+ - n_-)/2 = n_+. 
$

Using $n_M(\rm BI)=0$ as a reference, we obtain the mirror Chern number for the TI and TSC 
by calculating the change of $n_M$ across the phase transition to the BI. Due to the double counting of electrons and holes, the mirror Chern number of an insulator defined in Nambu space is twice the value of that defined previously\cite{teofukane}. As a result,  a direct transition from TI to BI at $\Delta=0$  changes $n_M$ by two. For $\Delta \neq 0$, this transition is split into two transitions with an intermediate TSC phase, so that each transition changes $n_M$ by one.  
Therefore we have
 \beq
 n_M({\rm TI}) = 2 n_M({\rm TSC}).  \label{mc}
 \eeq 
The fact that TI and TSC have mirror Chern numbers of the {\it same} sign will play a key role in 
our analysis of gapless excitations on their surfaces.  

{\bf Surface States}: Consider a semi-infinite  Bi$_2$Se$_3$ crystal occupying $z<0$,  which is naturally cleaved between QLs (see Fig.1). 
The boundary condition corresponding to such a termination in $k\cdot p$ theory is 
\beq
\sigma_z \psi(z=0) = \psi(z=0) \label{bc}.
\eeq
As explained in Ref\cite{fuberg}, this boundary condition reflects the vanishing of the wavefunction on the bottom layer ($\sigma_z=-1$) at $z=0$,   
whereas the other termination within the QL (not yet reported in experiments)  corresponds to a different boundary condition $\sigma_z\psi=-\psi$ at $z=0$\cite{notebc}.  

First we study surface states of Bi$_2$Se$_3$ by solving the differential equation $H(k_x, k_y, -i\partial_z)\psi=E \psi$ subject to the boundary condition (\ref{bc}). We obtain two sets of exact solutions $\psi_{\pm}(\bk_\parallel,z)$
with the energy-momentum dispersion $E_\pm(k) = \pm v k$.  
\beq
\psi_\pm(\bk_\parallel, z) &=& e^{z/l } (1, 0)_\sigma \otimes ( 1, \pm ie^{i \phi})_s,  \label{sf}
\eeq
where $ l = - v_z/m$ is the decay length; $\phi$ is the azimuthal angle of $\bk_\parallel$; the subscripts $\sigma$ and $s$ denote the orbital $\sigma_z$ and spin $s_z$ basis. 
In order for the surface states to exist in a TI, we must have decaying solutions in the $-z$ direction, which implies $v_z m<0$.  
The spin polarization of $\psi_\pm$ is locked to its momentum, forming a  two-dimensional Dirac cone. In contrast, a BI has $v_z m >0$ 
for which no surface state solutions exist. 

%
Next we study surface Andreev bound states of Cu$_x$Bi$_2$Se$_3$. 
 We start by solving the BdG equation at $\bk_\parallel=0$. A Kramers pair of zero-energy eigenstates $\psi^0_{\alpha=\pm}(z)$ with mirror eigenvalues $\tilde{M}= i \cdot \alpha$ is expected from the topology and  symmetry of $H_{\rm BdG}(\bk_\parallel=0)$. $\psi^0_{\alpha}(z)$ satisfies a reduced $4$-component equation:
\beq
 [ (m \sigma_x -i v_z \sigma_y \partial_z   - \mu) \tau_z + \Delta \sigma_y \tau_x ] \psi^0_\alpha(z) =0 .  \label{sfeq1}
\eeq 
By multiplying both sides of Eq.(\ref{sfeq1}) with $\tau_z$, it becomes clear that $\psi^0_\alpha$ is an eigenstate of $\tau_y$. 
The corresponding eigenvalue is given by ${\rm sgn }(v_z) $ in order to have a decaying solution. 
Eq.(\ref{sfeq1}) then reduces to a two-component equation in orbital space, which has two independent solutions:
\beq
\xi_\pm(z) = (1, e^{\pm i \theta})_\sigma \cdot e^{\pm i k_{F} z + \kappa z}. \label{xi}
\eeq
Here $k_{F}$ is the Fermi momentum in the $z$ direction, given by $k_{F} = \sqrt{\mu^2 - m^2 }/v_z $;
$\kappa$ is the inverse decay length  given by $\kappa = \Delta/|v_z|>0$; 
$\theta$  is an angle defined by $e^{i \theta} = (m + i \sqrt{\mu^2 - m^2})/\mu$. 
We now choose a suitable linear combination of $\xi_+$ and $\xi_-$ to satisfy the boundary condition (\ref{bc}) and obtain 
the wavefunction of SABS: 
\beq
\psi^0_{\alpha}(z) &\propto& e^{\kappa z}   (\sin(k_{F} z  - \theta), \sin(k_{F} z))_\sigma \otimes  \nonumber \\
& & \left[ (1, -\alpha)_s \otimes (1, 0)_\tau + i {\rm sgn}(v_z) (1, \alpha)_s \otimes (0, 1)_\tau  \right], \nonumber 
\eeq
where $\tau=\pm 1$ denote the particle/hole basis. 
$\psi^0_{s}(z)$ is  particle-hole symmetric: $\Xi \psi^0_{s}(z) = \psi^0_s(z)$ up to an overall phase ($\Xi=s_y \tau_y K$), and therefore 
represents a Majorana fermion.  
In the limit $m=0$, $\psi^0_\alpha$ agrees with the result from Ref.\cite{fuberg}. 

Away from $k=0$, the Kramers doublet $\psi^0_{+}$ and $\psi^0_{-}$ is split by spin-orbit coupling term in (\ref{kp}).  
To lowest order in $k$, the SABS dispersion $\epsilon_\alpha(k)$  is linear and given by 
$ \epsilon_\alpha(k) =  \alpha \tilde{v} k $. 
The velocity $\tilde{v}$ is obtained from first-order perturbation theory:
\beq
\tilde{v} &=& v \frac{\Delta^2 + {\rm sgn}(v_z)\Delta m}{\Delta^2 + {\rm sgn}(v_z) \Delta m + \mu^2} \label{velocity}
\eeq
Since $\Delta \ll |m| <\mu$  in weak-coupling superconductors (provided $m$ is not vanishingly small), 
 (\ref{velocity}) simplifies to 
 \beq
 \tilde{v} \simeq v \cdot {\rm sgn}(v_z )\Delta m / \mu^2. 
 \label{v}
 \eeq 

While the SABS in previous studies of {\it single-band} superconductors can often be obtained by quasi-classical methods 
from the geometry of the Fermi surface alone, this is not the case for {\it multi-orbital} systems such as Cu$_x$Bi$_2$Se$_3$.  
Here both  the wavefunction and dispersion of SABS depend on the dimensionless factor $m/\mu$, which arises from the orbital character of electron wavefunctions on the Fermi surface.

%
To deduce the behavior of SABS dispersion for large $k$,  
we make use of the mirror Chern number introduced earlier. 
According to the principle of bulk-boundary correspondence, the sign of mirror Chern number $n_M$ determines the helicity of surface states as follows.  
$n_M < 0$ implies that the branch of surface states at $k_x=0$ with mirror eigenvalues $\mp i$ (i.e., $s_x=\pm1$) moves (anti-)clockwise with respect to $+x$ axis at the edge of the $yz$ plane, which means that the electron's spin and angular momentum from circulation at the edge are parallel.   
 Such a helicity forces the dispersion of the $s_x=+1$ surface band to eventually merge into the $E > 0$ bulk quasiparticle continuum at a large positive $k$. 
 
Given this correspondence between the mirror Chern number and the helicity, the  SABS in the TSC must have the same helicity as the TI surface states, because the mirror Chern numbers  (\ref{mc}) of the two phases have the same sign.  
 Moreover, the helicity of   Bi$_2$Se$_3$ surface states is directly given by the sign of Dirac velocity $v$ (since the dispersion is monotonic).  
Comparing the SABS velocity $\tilde{v}$ in (\ref{v}) with $v$, we find that  
${\rm sgn}(\tilde{v})= {\rm sgn}(v)$ for  TSC in a doped BI ($v_zm>0$),  whereas 
 ${\rm sgn}(\tilde{v})= -{\rm sgn}(v)$ for TSC in a doped TI ($v_zm<0$).  The latter case deserves further attention.   
In order to fulfill the helicity requirement, the two branches of SABS with opposite mirror eigenvalues must eventually switch places and thus cross $E=0$ once again at some finite $k$ (or an odd number of times). 
As long as perturbations do not change the sign of $\tilde v$ at $k=0$, the second crossing is robust due to the mirror Chern number argument. 
Similar reasoning was previously applied to surface states in TIs\cite{teofukane, antimony, murakami}. 

To demonstrate the SABS explicitly, we construct a two-orbital tight-binding model in the rhombohedral lattice 
shown in Fig.1 and calculate the SABS dispersion numerically. The Hamiltonian is defined as follows: 
\beq
H=H_{0}+H_{12}+ H_{\rm soc} + H'_{12} . 
\eeq
$
H_{0}= \sum_{<ij>}  t_{0}c_{i\alpha}^{\dagger}c_{j\alpha} \nonumber 
$
and 
$
H_{12}=\sum_{<i\in1,j\in2>}t_{1}c_{i\alpha}^{\dagger}c_{j\alpha} + \sum_{<i\in1,j'\in2>} 
t_{2} c_{i\alpha}^{\dagger}c_{j'\alpha} 
$
describes  spin-independent nearest neighbor hopping within the same layer ($t_0$), as well as 
between two neighboring layers within a QL ($t_1$) and on two adjacent QLs ($t_2$).   
$H_{\rm soc}$ describes the Rashba-type spin-orbit coupling associated with nearest-neighbor intra-layer hopping, which take opposite signs on the top and bottom layer of the unit cell due to opposite local electric fields along the $z$ direction: 
\beq
H_{\rm soc}&=&(\sum_{<ij>\in 1} - \sum_{<ij>\in 2}) \frac{i \lambda}{2}c_{i\alpha}^{\dagger}{\vec s}_{\alpha\beta}c_{j\beta}\cdot(\hat{z}\times \mathbf{a}_{ij}), \nonumber
\eeq
 where $\mathbf{a}_{ij}=\frac{1}{2}\epsilon_{ijk}(\mathbf{R}_j-\mathbf{R}_k)$ denote the vectors joining nearest neighbors within a layer, and $\mathbf{R}_{1,2,3}$ are the Bravais lattice vectors. The last term $H_{12}'$ describes inter-layer second nearest neighbor ($t_{3})$ hopping within a QL:
$
H'_{12}= \sum_{<<i\in1,j\in2>>}t_{3}c_{i\alpha}^{\dagger}c_{j\alpha} + h.c. \nonumber
$
 
This tight-binding model is constructed to reproduce the $k\cdot p$ Hamiltonian (1) of Bi$_2$Se$_3$ in the small $k$ limit as follows: 
$m=3(t_1+t_2+t_3)$, $v_z=3t_2 c$, and $v=\frac{9}{2}\lambda a^2$, where $a=|\mathbf{a}_{ij}|$ and $c=|\frac{1}{3}(\mathbf{R}_1+\mathbf{R}_2+\mathbf{R}_3)|$. 
Moreover, since there is hopping beyond two neighboring layers in $H$, the open boundary condition in this lattice model correctly reproduces the desired boundary condition (\ref{bc}) in the continuum theory\cite{anton}.   
It is to be emphasized that  this tight-binding model does {\it not} aim to 
describe the band structure of Bi$_2$Se$_3$ in the entire Brillouin zone\cite{vanderbilt}. 

\begin{figure}
\centering
\includegraphics[width=3.5in]{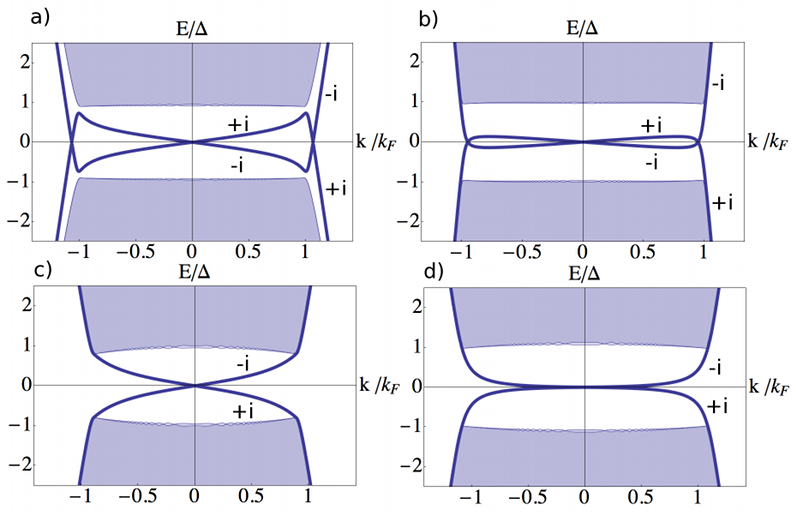}
\caption{SABS dispersion for the tight-binding model in which a) $m=-0.3<0$, $\mu_1=0.6$ and b) $m=-0.3<0$, $\mu_2=1$ correspond to doped TI;  
c) $m=0.3>0$, $\mu_1=0.6$ corresponds to a doped BI; d) $m=0$, $\mu_1=0.6$ corresponds to a doped zero-gap semiconductor.  The mirror eigenvalues are displayed near each branch of SABS, Note that in a doped TI, SABS is twisted with a second crossing near Fermi momentum.}
\end{figure}

The bulk and surface bands of the above tight-binding model are displayed in Figure 1b. To include superconductivity, we add the following odd-parity pairing term in the Hamiltonian:
\beq
H_{\rm MF} = H + \sum_{<i\in 1,j\in 2>} \frac{\Delta}{2} (c^\dagger_{i\uparrow}c^\dagger_{j\downarrow} + c^\dagger_{i\downarrow}c^\dagger_{j\uparrow} ) + h.c . \nonumber
\eeq 
Fig.3 shows the resulting SABS dispersion in both doped TI ($m<0$, Fig.3a-b) and doped BI ($m>0$, Fig.3c). 
The SABS in both cases have a branch of linearly dispersing Majorana fermion at $k=0$, which signifies a three-dimensional topological superconductor. 
Moreover, the SABS in doped TI has a ``twisted'' dispersion with a second crossing near Fermi momentum, as anticipated from our earlier mirror Chern number argument.

The tight-binding calculation sheds light on the origin of twisted SABS. The presence of the second crossing is a remnant of the TI surface states in the normal state. 
This can be easily understood in  the case where electron surface states remain well-defined at the Fermi energy in the normal state, e.g, at the chemical potential $\mu_1$ in Fig.1c. 
The surface states at $\bk$ and $-\bk$ have {\it opposite} mirror eigenvalues, whereas $\Delta$ only pairs
states with the {\it same} mirror eigenvalues. Due to this symmetry incompatibility, surface states must remain gapless even in the presence of such an 
odd-parity pairing. Therefore, the Majorana fermion SABS at $k=0$ smoothly evolves into TI surface states near Fermi momentum. 
Indeed we find that the decay length of SABS at and beyond the second crossing is comparable to that of the electron surface state given by $|v_z|/|m|$, 
which is much shorter than the typical decay length of SABS given by $|v_z|/\Delta$.  This is further confirmed by an analytic calculation 
using the $k\cdot p$ BdG Hamiltonian (\ref{bdg}). We search for  an $E=0$ solution at a nonzero $k_0$ and find $k_0$ is the nontrivial solution of 
\beq
2{\rm Re}(W)+\frac{m}{E_F}(-1+|W|^2)=0, 
\eeq where
\beq
W\equiv \frac{vk_0-i(\Delta+iE_F)}{\sqrt{(vk_0)^2+(\Delta+iE_F)^2}}.
\eeq  
For $\Delta = 0$ and $m<0$, the solution is given by $k_0= \mu/v$, which  represents the TI surface state at Fermi energy. 
For $\Delta \neq 0$, the solution is perturbed: $k_0= \frac{\mu}{v}(1-\frac{\Delta^2}{2m^2})$ to leading order in  $\Delta$.  No solution exists for $m>0$. 

What happens if  there are {\it no} surface states at the Fermi energy in the normal state of a doped TI? 
For example, at the chemical potential $\mu_2$ in Fig.1c, the TI surface states have merged into the bulk\cite{merge}.  
Remarkably,  after turning on the odd-parity pairing (Fig.3b), the SABS still has the second crossing, as required by the mirror Chern number. 
The resulting gapless SABS near Fermi momentum have significant particle-hole mixing and therefore cannot be interpreted as unpaired TI surface states.   
This represents a new type of surface Andreev bound state, which originates from the interplay between 
band structure and unconventional superconductivity. 
Such SABS defy a conventional quasi-classical description and are worth further study.

Finally, we relate our findings to the recent point-contact spectroscopy experiment on Cu$_x$Bi$_2$Se$_3$\cite{ando}. A zero-bias conductance 
peak with a width of $0.6$mV was clearly observed below $1.2$K, which indicates the existence of SABS.     
A natural candidate is SABS of the fully-gapped odd-parity topological superconductor phase proposed in Ref.\cite{fuberg}. 
In this Letter, we found that the SABS of TSC in a doped TI have unusual and topologically robust features. As shown in Fig.3a and b, the second crossings of SABS at finite $k$ along all in-plane directions make the density of states at zero energy finite, in sharp contrast to the vanishing density of states at $k = 0$ characteristic of the linear dispersion in 2D as found in a  doped BI (Fig. 3c). 
Moreover, for small $m/\mu$ the SABS dispersion is quite flat as shown in Fig.3b and 3d.  
These features may naturally explain the experimental observations, and will be studied in a future work. 

While the main focus of this Letter is Cu$_x$Bi$_2$Se$_3$, we end by discussing the applicability of our findings to a large class of superconducting doped semiconductors with inversion symmetry. Potential candidates include Bi$_2$Te$_3$\cite{bite} and TlBiTe$_2$\cite{thallium} under pressure, PbTe\cite{pbte}, SnTe\cite{snte}, and GeTe\cite{gete}. Provided that the Fermi surface is centered at time-reversal-invariant momenta, the Dirac-type relativistic $k\cdot p$ Hamiltonian gives a valid low-energy description of its band structure\cite{fukane}. Moreover, if the pairing is odd under spatial inversion and fully gapped, the system is (almost) guaranteed to be a topological superconductor according to our criterion\cite{fuberg, qifu}. In that case, our findings of 
unusual surface Andreev bound states apply directly. In addition, if the pairing symmetry of noncentrosymmetric superconductors (such as YPtBi\cite{yptbi}) has a dominant odd-parity component, their properties are often inherited from the centrosymmetric limit.  We hope this work will encourage further explorations of unconventional surface Andreev bound states and stimulate the search for topological superconductors.  

{\it Note:}  Two recent studies \cite{haolee, ando} have calculated the surface spectral function numerically in Cu$_x$Bi$_2$Se$_3$ tight-binding models different from ours. The connection to our work on Majorana fermion surface Andreev bound state remains to be understood.







{\it Acknowledgement}: We thank Erez Berg and Yang Qi for helpful discussions, as well as Yoichi Ando, Anton Akhmerov, and David Vanderbilt for helpful comments on the manuscript. TH is supported by the U.S. Department of Energy under cooperative research agreement Contract Number DE-FG02-05ER41360 and the National Science Foundation Graduate Research Fellowship under Grant No. 0645960.
LF would like to thank Physics Department at MIT, Institute of Physics in China, and Institute of Advanced Study at Tsinghua University for generous hosting, as well as 
 the Harvard Society of Fellows for support.


\begin{thebibliography}{10}
\bibitem{notebc} In contrast,  the boundary condition in Ref.\cite{liu}  does not differentiate the two surface terminations.
\bibitem{merge}  Surface states  merge into the bulk at some large $k$ for which the $k.p$ approximation is no longer valid.
\end{thebibliography}
\end{document}